\documentclass[aps,prd,onecolumn,11pt,superscriptaddress,nofootinbib,floatfix]{revtex4-2}
\usepackage[T1]{fontenc}
\usepackage[utf8]{inputenc}
\usepackage{lmodern}
\usepackage{amsmath,amssymb,amsthm}
\usepackage{bm}
\usepackage{xcolor}
\usepackage[colorlinks=true,linkcolor=blue,citecolor=blue,urlcolor=blue]{hyperref}

\theoremstyle{plain}
\newtheorem{theorem}{Theorem}
\newtheorem{proposition}{Proposition}
\newtheorem{lemma}{Lemma}
\theoremstyle{definition}
\newtheorem{definition}{Definition}
\newtheorem{remark}{Remark}

\begin{document}

\title{Einstein--Yang--Mills Plane Waves: Uniqueness and an Exact Curvature Double Copy}

\author{Metin G{\"u}rses}
\email{gurses@fen.bilkent.edu.tr}
\affiliation{Department of Mathematics, Bilkent University, 06800 Ankara, T{\"u}rkiye}
\author{Tahsin \c{C}a{\u g}r{\i} \c{S}i\c{s}man}
\email{tahsin.c.sisman@gmail.com}
\affiliation{Institute of Nuclear Sciences, Hacettepe University, 06800 Ankara, T{\"u}rkiye}
\author{Bayram Tekin}
\email{bayram.tekin@bilkent.edu.tr}
\affiliation{Department of Physics, Bilkent University, 06800 Ankara, T{\"u}rkiye}

\begin{abstract}
Coleman found plane waves of Yang--Mills theory that are genuinely non-abelian in four dimensions and yet obey linear field equations, and G\"uven showed that they remain exact when gravity is included. We study these waves in $D$ dimensions. Coleman's profile, which is linear in the transverse coordinates, is singled out twice: up to gauge, it is the only profile with bounded energy density, and the only one for which the full Riemann tensor equals the color-contracted square of the Yang--Mills field strength that sources it. In anti-de Sitter space, we construct a closed-form family for every $D\geq4$; there, the full Riemann tensor cannot equal this color square. Waves whose transverse space is a sphere are always singular. In quadratic and Gauss--Bonnet gravity, a hyperbolic transverse space avoids this only if the waves defocus light or the higher-curvature terms become comparable to the Einstein term. All solutions extend, with changed coefficients, to generic gravity and gauge theories.
\end{abstract}

\maketitle
\enlargethispage{-\baselineskip} 

\section{Introduction}

Exact solutions of non-abelian gauge theory that describe radiation are rare because the Yang--Mills equations are nonlinear. Coleman observed \cite{col1} that the nonlinear terms drop out identically when the potential points along a covariantly constant null direction and does not depend on the conjugate null coordinate. The resulting plane waves can nevertheless be genuinely non-abelian: their field-strength components need not commute. More general null Yang--Mills waves were constructed in \cite{campbell-morgan,trautman-null}, and G\"uven \cite{guv1} showed that the gravitational field of Coleman's waves can be included exactly. A supersymmetric generalization of Coleman's waves was given by Dereli and G\"uven \cite{dereli-guven}. The spacetime is a pp-wave whose metric profile is the color-contracted square of the gauge profile, $V\propto\gamma_{ab}A^aA^b$. This is a literal version of the slogan ``gravity $=$ (gauge)$^2$''. That slogan underlies the double copy: the observation that gravitational scattering amplitudes \cite{bern,bcj2010}, and some exact gravitational solutions \cite{mon,mar,godoy}, can be obtained from gauge-theory ones by replacing color data with a second copy of kinematic data. Coleman's waves were later embedded in Kundt spacetimes of Petrov type III \cite{fuster-vh,fuster}. Non-abelian pp-waves in anti-de Sitter (AdS) space were found in gauged supergravity by Cariglia, Gibbons, G\"uven and Pope \cite{cggp}, who also gave both the Yang--Mills equation and the sourced gravitational equation in AdS$_D$, and Gleiser and Dotti \cite{gleiser-dotti} constructed Lovelock--Yang--Mills waves on (A)dS backgrounds in every dimension. For null gauge fields on Kerr--Schild--Kundt backgrounds, which include waves in AdS, the Einstein--Yang--Mills equations were reduced to linear equations in \cite{gur-tek}.

Two properties make these solutions insensitive to corrections. (a) The spacetimes considered here, namely pp-waves, Kerr--Schild--Kundt metrics, and pp-waves over $\mathbb R^{1,1}\times S^{D-2}$, are \emph{almost universal} \cite{gur1,gur2,gur3,ghst,gur4,gur5,smooth,gur6,gur-sis-tek,ort4}. That is: for every metric theory of gravity whose Lagrangian is built from the curvature and its derivatives, the field equations on these spacetimes reduce to a single linear equation for the profile $V$, and the theory enters only through a few constants. (b) The gauge fields are \emph{null}: all their scalar invariants vanish. As a consequence, any gauge Lagrangian built algebraically from the field strength reduces to the Yang--Mills one on them (Sec.~\ref{sec:gauge-univ}; see \cite{ort1,ort2,ort3,hop2018} for electrodynamics). For null Yang--Mills fields, the general result for algebraic corrections was proved by Kuchynka \cite{kuchynka}. G\"uven \cite{guv1987} earlier studied higher-order corrections to Yang--Mills plane waves in ten-dimensional supergravity and argued for their all-orders string exactness. The solutions are therefore exact in a very large class of gravity and gauge theories.

Our results are as follows.
\begin{enumerate}
\item[(i)] In $D$-dimensional pp-wave spacetimes, Coleman's profile, which is linear in the transverse coordinates, is, up to gauge, the \emph{only} gauge profile with bounded energy density (Theorem~\ref{thm:unique}). In $D=4$ this was pointed out by G\"uven \cite{guv1}; see also \cite{fuster-vh,fuster}. We give a short proof valid in every dimension. The $D$-dimensional Coleman--G\"uven solution itself is a special case of the solutions of \cite{gur-tek}.
\item[(ii)] For Coleman's profile, and only for it, the full Riemann tensor equals $\kappa\gamma_{ab}F^a_{\mu\nu}F^b_{\rho\sigma}$ (Theorem~\ref{thm:dc}). In four dimensions, this is a Weyl double copy with a constant ``zeroth copy'' in which the gauge field is the source of the spacetime rather than a test field.
\item[(iii)] For a Coleman wave with non-commuting amplitudes, the field strength is as symmetric as the geometry, but the potential is not, even up to gauge (Proposition~\ref{prop:nosym}). Colored test particles can detect this difference; neutral ones do not.
\item[(iv)] The reduced equations for non-abelian waves in AdS$_D$ are known \cite{cggp,gur-tek}. We construct a closed-form family in every dimension $D\geq4$ (Proposition~\ref{prop:ads}). The identity of (ii) is incompatible with the nonzero AdS scalar curvature. In $D=4$, even the Weyl parts are not generically proportional.
\item[(v)] For pp-waves over $\mathbb R^{1,1}\times S^{D-2}$, every non-trivial solution is singular (Proposition~\ref{prop:noreg}). A negatively curved transverse space would avoid this. In quadratic and Gauss--Bonnet gravity, however, it requires either waves that defocus light rays or curvatures at which the higher-curvature terms become comparable to the Einstein term (Theorem~\ref{thm:hyp}).
\item[(vi)] All of the above carry over to generic gravity and gauge theories, with explicitly modified coefficients (Sec.~\ref{sec:generic}). For the gauge sector, this follows from \cite{kuchynka}.
\end{enumerate}

The paper is organized as follows. Section~\ref{sec:pp} treats flat transverse space and proves (i), and Sec.~\ref{sec:dc} proves (ii). Section~\ref{sec:sym} explains both results through the symmetry of the wave fronts and discusses test particles. Section~\ref{sec:ads} constructs the AdS waves, Sec.~\ref{sec:sphere} treats spherical transverse spaces, and Sec.~\ref{sec:generic} extends everything to generic theories. We conclude in Sec.~\ref{sec:concl}.

\emph{Conventions.} The signature is $(-,+,\dots,+)$. The field equations are written as $G_{\mu\nu}+\Lambda g_{\mu\nu}=\kappa T_{\mu\nu}$, and the gauge coupling is set to one. Greek indices are spacetime indices, $i,j,k$ label transverse coordinates, and $a,b,c$ label the Lie algebra. In AdS, $m,m'$ label the coordinates $y^m$. A dot or a prime denotes $d/du$, except in Sec.~\ref{sec:geodesics}, where dots on particle coordinates denote $d/d\tau$ and primes denote $d/du$. Every solution stated in this paper was checked against the full field equations by computer algebra, for the gauge group $SU(2)$ and several dimensions.

\section{Non-abelian plane waves with flat transverse space}\label{sec:pp}

\subsection{Gauge-theory conventions}

Let $G$ be a compact semi-simple Lie group with Lie algebra $\mathfrak g$, generators $T_a$ ($a=1,\dots,\dim\mathfrak g$), and real structure constants, $[T_b,T_c]=f^a{}_{bc}T_a$. A Lie-algebra-valued potential $A_\mu=A^a_\mu T_a$ has field strength and covariant derivative
\begin{equation}
F^a_{\mu\nu}=\partial_\mu A^a_\nu-\partial_\nu A^a_\mu+f^a{}_{bc}A^b_\mu A^c_\nu ,\qquad
(D_\mu X)^a=\nabla_\mu X^a+f^a{}_{bc}A^b_\mu X^c .
\end{equation}
The Yang--Mills equations and the energy--momentum tensor are
\begin{equation}
(D_\mu F^{\mu\nu})^a=0,\qquad
T^{\rm YM}_{\mu\nu}=\gamma_{ab}\Big(F^a{}_{\mu}{}^{\alpha}F^b{}_{\nu\alpha}-\frac14 g_{\mu\nu}F^a_{\alpha\beta}F^{b\,\alpha\beta}\Big),
\label{YM}
\end{equation}
where $\gamma_{ab}\equiv-f^c{}_{ad}f^d{}_{bc}$ is minus the Killing form of $\mathfrak g$. For compact semi-simple $G$, it is positive definite and invariant under the adjoint action; for $su(2)$ with $f^a{}_{bc}=\epsilon_{abc}$, $\gamma_{ab}=2\delta_{ab}$. Positivity of $\gamma_{ab}$ is what makes the energy density of the waves below non-negative.

Under a gauge transformation $\mathcal U(x)\in G$ the field strength transforms by conjugation, $F_{\mu\nu}\to \mathcal U^{-1}F_{\mu\nu}\mathcal U$, at each point. This gives a simple test of non-abelianness.

\begin{definition}
A configuration is called \emph{abelian} if there exists a gauge in which its potential $A_\mu(x)$ takes values in one abelian subalgebra of $\mathfrak g$, the same for all $x$ and $\mu$.
\end{definition}

In an abelian configuration all commutators $[F_{\mu\nu},F_{\rho\sigma}]$ vanish at every point. Conjugation preserves the vanishing of a commutator, so a single point where $[F_{\mu\nu},F_{\rho\sigma}]\neq0$ proves that a configuration is not abelian. This criterion is sufficient but not necessary: pointwise commutativity of the field-strength components does not by itself imply that a configuration is abelian.

\subsection{Null gauge fields}

All gauge fields that we shall study are of the following form, on several different spacetimes.

\begin{lemma}\label{lem:null}
Let $u$ be a function whose gradient $\lambda_\mu=\partial_\mu u$ is null, $\lambda^\mu\lambda_\mu=0$, and let
\begin{equation}
A^a_\mu=A^a\,\lambda_\mu,\qquad \lambda^\mu\partial_\mu A^a=0 .
\label{ansatz}
\end{equation}
Then:
\begin{enumerate}
\item[(a)] $F^a_{\mu\nu}=\lambda_\nu\partial_\mu A^a-\lambda_\mu\partial_\nu A^a$, and $\lambda^\mu F^a_{\mu\nu}=0$;
\item[(b)] the non-abelian term in the Yang--Mills equations vanishes, and
\begin{equation}
(D^\mu F_{\mu\nu})^a=\nabla^\mu F^a_{\mu\nu}=\lambda_\nu\Box A^a+\partial^\mu A^a\,(\nabla_\mu\lambda_\nu+\nabla_\nu\lambda_\mu)-(\nabla\!\cdot\!\lambda)\,\partial_\nu A^a ;
\label{divF}
\end{equation}
\item[(c)] $F^a_{\alpha\beta}F^{b\,\alpha\beta}=0$, $F^a{}_\mu{}^\alpha F^b{}_{\nu\alpha}=(\partial A^a\!\cdot\!\partial A^b)\lambda_\mu\lambda_\nu$, and hence
\begin{equation}
T^{\rm YM}_{\mu\nu}=\lambda_\mu\lambda_\nu\gamma_{ab}\partial A^a\!\cdot\!\partial A^b ,\qquad \partial A^a\!\cdot\!\partial A^b\equiv g^{\alpha\beta}\partial_\alpha A^a\partial_\beta A^b .
\label{Tnull}
\end{equation}
The quantity $\gamma_{ab}\,\partial A^a\!\cdot\!\partial A^b$ is non-negative.
\end{enumerate}
\end{lemma}

\begin{proof}
(a) Since $d\lambda=0$, $F^a=dA^a\wedge\lambda$, and the quadratic term $f^a{}_{bc}A^bA^c\lambda_\mu\lambda_\nu$ vanishes by antisymmetry of $f^a{}_{bc}$. The contraction $\lambda^\mu F^a_{\mu\nu}$ vanishes because $\lambda^2=0$ and $\lambda\cdot\partial A^a=0$. (b) The non-abelian term $f^a{}_{bc}A^b\lambda_\mu F^{c\,\mu\nu}$ vanishes by (a). For (\ref{divF}), use $\lambda^\mu\nabla_\mu\partial_\nu A^a=\nabla_\nu(\lambda\cdot\partial A^a)-(\nabla_\nu\lambda^\mu)\partial_\mu A^a=-(\nabla_\nu\lambda^\mu)\partial_\mu A^a$. (c) Expand the products and use $\lambda^2=0$ and $\lambda\cdot\partial A^a=0$. Finally, $\partial A^a$ is orthogonal to the null vector $\lambda$, so it is spacelike or null; since $\gamma_{ab}$ is positive definite, $\gamma_{ab}\,\partial A^a\cdot\partial A^b\geq0$.
\end{proof}

Thus, for every background below, the Yang--Mills equations are \emph{linear} in $A^a$ and the stress tensor is that of null dust, i.e.\ of radiation moving at the speed of light along $\lambda$.

\subsection{Coleman's waves in flat spacetime}

Take $D$-dimensional Minkowski spacetime in null coordinates $x^\mu=(u,v,y^i)$, $i=1,\dots,n$, $n\equiv D-2$:
\begin{equation}
ds^2=2\,du\,dv+\delta_{ij}dy^idy^j ,\qquad \lambda_\mu dx^\mu=du,\qquad \lambda^\mu\partial_\mu=\partial_v .
\end{equation}
The ansatz (\ref{ansatz}) is Coleman's \cite{col1}: $A^a=A^a(u,\vec y)$ is independent of $v$. Since $\nabla_\mu\lambda_\nu=0$, Eq.~(\ref{divF}) gives $\nabla^\mu F^a_{\mu\nu}=(\nabla^2A^a)\lambda_\nu$, where $\nabla^2\equiv\delta^{ij}\partial_i\partial_j$ is the transverse Laplacian. The Yang--Mills equations reduce to
\begin{equation}
\nabla^2A^a=0 .
\end{equation}
For each $u$ the gauge profile is an arbitrary harmonic function of the transverse coordinates. In $D=4$ it can locally be written as the real part of a holomorphic function:
\begin{equation}
A^a(u,y^1,y^2)=\mathrm{Re}\,\Phi^a(u,\zeta),\qquad \partial_{\bar\zeta}\Phi^a(u,\zeta)=0,\qquad \zeta\equiv y^1+iy^2,\qquad \partial_{\bar\zeta}\equiv\frac12\left(\partial_{y^1}+i\partial_{y^2}\right).
\end{equation}

The non-vanishing field-strength components are $F^a_{ui}=-\partial_iA^a$, and
\begin{equation}
[F_{ui},F_{uj}]^a=f^a{}_{bc}\,\partial_iA^b\,\partial_jA^c .
\end{equation}
Coleman's own solution is the profile linear in $\vec y$,
\begin{equation}
A^a=p^a{}_i(u)\,y^i ,
\label{coleman_A}
\end{equation}
where we call $p^a{}_i(u)$ the color amplitudes and write $p_i\equiv p^a{}_iT_a$. By the test above, it is non-abelian whenever $f^a{}_{bc}p^b{}_i(u)p^c{}_j(u)\neq0$ for some $i,j$ and $u$, which requires two transverse directions, i.e.\ $D\geq4$. In $D=3$, there is only one independent field-strength component, so all field-strength components commute at each point, and this test cannot decide whether the configuration is non-abelian. We restrict the subsequent discussion to $D\geq4$.

\subsection{Gravitating non-abelian waves in \texorpdfstring{$D$}{D} dimensions}

G\"uven \cite{guv1} coupled Coleman's waves to gravity in $D=4$. In $D$ dimensions we take the pp-wave \cite{Kundt,Stephani_et_al}
\begin{equation}
ds^2=2\,du\,dv+2V(u,\vec y)\,du^2+\delta_{ij}dy^idy^j ,
\qquad g_{\mu\nu}=\eta_{\mu\nu}+2V\lambda_\mu\lambda_\nu .
\label{ppwave}
\end{equation}
The coordinates in (\ref{ppwave}) are called Brinkmann coordinates \cite{brinkmann}. The vector $\lambda_\mu dx^\mu=du$ remains null and covariantly constant, the Ricci scalar vanishes, and
\begin{equation}
G_{\mu\nu}=R_{\mu\nu}=-\lambda_\mu\lambda_\nu\nabla^2V .
\end{equation}
Lemma~\ref{lem:null} applies unchanged. Since $\sqrt{-g}=1$ and $g^{\mu\nu}=\eta^{\mu\nu}-2V\lambda^\mu\lambda^\nu$, $\Box A^a=\nabla^2A^a$ for $v$-independent $A^a$, and $\partial A^a\cdot\partial A^b=\nabla A^a\cdot\nabla A^b\equiv\delta^{ij}\partial_iA^a\partial_jA^b$. The Einstein--Yang--Mills equations $G_{\mu\nu}=\kappa T^{\rm YM}_{\mu\nu}$, $(D_\mu F^{\mu\nu})^a=0$ therefore reduce to
\begin{equation}
\nabla^2A^a=0,\qquad \nabla^2V=-\kappa\,\gamma_{ab}\,\nabla A^a\!\cdot\!\nabla A^b .
\label{ppred}
\end{equation}
For harmonic $A^a$ one has $\nabla^2(\gamma_{ab}A^aA^b)=2\gamma_{ab}\nabla A^a\cdot\nabla A^b$, so the general solution is
\begin{equation}
V=V_0-\frac{\kappa}{2}\,\gamma_{ab}A^aA^b,\qquad \nabla^2V_0=0 .
\label{guven}
\end{equation}
Here $V_0(u,\vec y)$ is an arbitrary transverse harmonic function: it is an independent vacuum gravitational wave superposed on the one produced by the gauge field. Terms in $V_0$ of the form $\beta_0(u)+\beta_i(u)y^i$ can be removed by a coordinate transformation.

\emph{Energy density.} An observer with unit timelike velocity $U^\mu$ measures
\begin{equation}
\mathcal E=T_{\mu\nu}U^\mu U^\nu=(\lambda_\mu U^\mu)^2\,\gamma_{ab}\nabla A^a\!\cdot\!\nabla A^b\ \geq0 .
\end{equation}
The factor $(\lambda\cdot U)^2$ is the usual Doppler factor of a null wave. We use observers with $(\lambda\cdot U)^2=1$, e.g.\ $U=\partial_u-(V+\tfrac12)\partial_v$.

\emph{The Coleman--G\"uven plane wave.} For the profile (\ref{coleman_A}), define the symmetric matrix of color-contracted amplitudes and its trace by
\begin{equation}
E_{ij}(u)\equiv\gamma_{ab}p^a{}_i(u)p^b{}_j(u),\qquad E(u)\equiv\delta^{ij}E_{ij}(u),\qquad i,j=1,\ldots,D-2.
\label{coleman_E}
\end{equation}
For the observers normalized above, the energy density $\mathcal E=E(u)$ is independent of $\vec y$ and vanishes only if all amplitudes vanish. If the amplitudes have compact support in $u$ (a \emph{sandwich wave}), the energy per unit transverse area, $\int\mathcal E\,du$, is finite. The total energy is infinite, as for any plane wave. Taking for $V_0$ the general quadratic vacuum profile, the complete solution in $D=4$ reads
\begin{equation}
\begin{aligned}
ds^2&=2\,du\,dv+dx^2+dy^2+2V\,du^2,\\
V&=K_+(u)(x^2-y^2)+2K_\times(u)\,xy-\frac{\kappa}{2}\gamma_{ab}\big[p^a{}_1(u)x+p^a{}_2(u)y\big]\big[p^b{}_1(u)x+p^b{}_2(u)y\big],\\
A&=\big[x p^a{}_1(u)+y p^a{}_2(u)\big]T_a\,du .
\end{aligned}
\label{box4}
\end{equation}
In $D$ dimensions, $xp^a{}_1(u)+yp^a{}_2(u)$ becomes $p^a{}_i(u)y^i$, and the first two terms of $V$ become $K_{ij}(u)y^iy^j$ with $K_{ij}$ symmetric and trace-free ($K_{11}=-K_{22}=K_+$, $K_{12}=K_\times$ in $D=4$). Setting $K_{ij}=0$ gives G\"uven's solution.

The tidal field of this wave is encoded in the curvature components
\begin{equation}
R_{uiuj}=-\partial_i\partial_jV=\kappa E_{ij}-2K_{ij},\qquad R_{uu}=\kappa\,E(u),\qquad
C_{uiuj}=\kappa\Big(E_{ij}-\frac{\delta_{ij}}{n}E(u)\Big)-2K_{ij},
\label{curvpp}
\end{equation}
where $C_{\mu\nu\rho\sigma}$ is the Weyl tensor; all components not related to these by symmetry vanish. For $K_{ij}=0$ the wave is conformally flat (pure null radiation, with no free gravitational wave) if and only if $E_{ij}\propto\delta_{ij}$. Conversely, the choice $2K_{ij}=\kappa(E_{ij}-\delta_{ij}E(u)/n)$ cancels the anisotropic tidal field of any color wave, so a suitable free gravitational wave makes any color wave conformally flat.

\emph{Polarizations.} Each color component of the gauge wave is a Maxwell plane wave whose transverse electric vector is $p^a{}_i(u)$. The wave therefore carries all $n=D-2$ polarizations of a gluon in each color direction. Because the Yang--Mills equations impose no condition on the $u$-dependence of the amplitudes, the waveform and the polarization state, whether linear, circular, or rotating in color space, are completely arbitrary. Gravity, on the other hand, sees the amplitudes only through the matrix $E_{ij}=\gamma_{ab}p^a{}_ip^b{}_j$. In an orthonormal color basis this is a sum, $E_{ij}=\sum_ap^a{}_ip^a{}_j$, of rank-one matrices, each of which is the stress of a single linearly polarized Maxwell wave. Gravity therefore sees a color wave exactly as it would see $\dim\mathfrak g$ independent Maxwell waves: the commutators $[p_i,p_j]$ do not enter $E_{ij}$. Through (\ref{curvpp}), the trace-free part of $E_{ij}$ sources the $D(D-3)/2$ independent components of the Weyl tensor, and its trace sources the Ricci part.

In four dimensions this can be made concrete. Write the two independent Weyl components as $\mathcal C_+\equiv C_{uxux}=-C_{uyuy}$ and $\mathcal C_\times\equiv C_{uxuy}$, and combine the color amplitudes into $\varpi^a\equiv p^a{}_1+i\,p^a{}_2$. For $K_{ij}=0$, Eq.~(\ref{curvpp}) gives
\begin{equation}
\mathcal C_++i\,\mathcal C_\times=\frac\kappa2\,\gamma_{ab}\,\varpi^a\varpi^b ,\qquad
\frac12R_{uu}=\frac\kappa2\,\gamma_{ab}\,\varpi^a\bar\varpi^b .
\label{helicity}
\end{equation}
Under a rotation of the wave front by an angle $\theta$, $\varpi^a$ is multiplied by a phase $e^{-i\theta}$; the Weyl part then picks up $e^{-2i\theta}$ (spin two), while the Ricci part is unchanged (spin zero).

The matrix $E_{ij}$ also obeys a bound. By the Cauchy--Schwarz inequality,
\begin{equation}
\Big(\frac12R_{uu}\Big)^2-\big(\mathcal C_+^2+\mathcal C_\times^2\big)=\kappa^2\det E\ \geq0 ,
\label{polbound}
\end{equation}
with equality if and only if $p_1$ and $p_2$ are parallel in the Lie algebra. For $E(u)>0$, define the normalized tidal anisotropy $\Pi\equiv\sqrt{\mathcal C_+^2+\mathcal C_\times^2}\,/\,\big(\tfrac12R_{uu}\big)$. With $K_{ij}=0$, it satisfies $0\le\Pi\le1$ and $\Pi^2=1-4\det E/E(u)^2$, comparing the Weyl tidal amplitude with the Ricci focusing. It equals one for a single Maxwell wave, and it is strictly smaller than one whenever $[p_1,p_2]\neq0$, since non-commuting amplitudes cannot be parallel. The converse holds only for gauge groups of rank one: for $SU(3)$, the amplitudes $p_1=c\,T_3$ and $p_2=c\,T_8$ commute, yet give $E_{ij}\propto\delta_{ij}$ and $\Pi=0$. For $su(2)$, where commuting elements are parallel, the deficit is exactly the commutator,
\begin{equation}
\Big(\frac12R_{uu}\Big)^2-\big(\mathcal C_+^2+\mathcal C_\times^2\big)=2\kappa^2\,\gamma_{ab}[p_1,p_2]^a[p_1,p_2]^b ,\qquad
\Pi^2=1-\frac{8\,\gamma_{ab}[p_1,p_2]^a[p_1,p_2]^b}{E(u)^2} .
\end{equation}
At the other extreme, $\Pi=0$ and the spacetime is conformally flat (Petrov type O) when $\gamma_{ab}p^a{}_1p^b{}_1=\gamma_{ab}p^a{}_2p^b{}_2$ and $\gamma_{ab}p^a{}_1p^b{}_2=0$, for example $p_1=p_0T_1$, $p_2=p_0T_2$ in $su(2)$, which is the maximally non-abelian configuration. A single Maxwell plane wave can never do this. In $D$ dimensions, conformal flatness requires $E_{ij}\propto\delta_{ij}$, hence $\operatorname{rank}E=n$ and $\dim\mathfrak g\geq D-2$. For $su(2)$ and $D\geq6$, a color wave alone ($K_{ij}=0$) therefore never gives a conformally flat spacetime.

\subsection{Uniqueness of Coleman's profile}

Coleman's profile has a field strength, and hence an energy density, that is uniform across the wave front. Could any other globally regular harmonic profile keep the energy density bounded? It cannot. This uniqueness was already noted in $D=4$ by G\"uven \cite{guv1}; see also \cite{fuster-vh,fuster}. The following proof applies in every dimension:

\begin{theorem}\label{thm:unique}
Let $\gamma_{ab}$ be positive definite and let $A^a(u,\cdot)\in C^2(\mathbb R^n)$ be harmonic for each $u$. Then $\mathcal E=\gamma_{ab}\nabla A^a\cdot\nabla A^b$ is bounded on $\mathbb R^n$ if and only if
\begin{equation}
A^a=p^a{}_i(u)\,y^i+c^a(u).
\end{equation}
The term $c^a(u)$ can be removed by a $u$-dependent gauge transformation, which preserves the ansatz (\ref{ansatz}).
\end{theorem}

\begin{proof}
If $A^a$ is affine in $\vec y$, $\mathcal E=E(u)$ is independent of $\vec y$. Conversely, let $\gamma_{\min}>0$ be the smallest eigenvalue of $\gamma_{ab}$. Then $\mathcal E\geq\gamma_{\min}\sum_{a,i}(\partial_iA^a)^2$, so each $\partial_iA^a$ is a bounded harmonic function on $\mathbb R^n$. By Liouville's theorem it is constant in $\vec y$. For the last statement, write $c=c^aT_a$ and let $\mathcal U(u)$ solve $\partial_u\mathcal U=-c\,\mathcal U$. Under $A_\mu\to \mathcal U^{-1}A_\mu \mathcal U+\mathcal U^{-1}\partial_\mu \mathcal U$, the only non-zero component $A_u=p_iy^i+c$ becomes $\mathcal U^{-1}p_i\mathcal U\,y^i$. The matrix $E_{ij}$ is invariant, because $\gamma_{ab}$ is.
\end{proof}

Physically, a non-abelian plane wave whose energy density is bounded over the entire wave front must be uniform. Every other harmonic profile either grows at infinity or is singular somewhere. Singular profiles are not unphysical: they describe the fields of sources, such as null color charges, placed at the singularities. We leave their analysis to future work.

\section{An exact double copy at the level of curvature}\label{sec:dc}

The metric profile (\ref{guven}) is quadratic in the gauge potential. To see what is special about this, recall the classical double copies. In the \emph{Kerr--Schild double copy} \cite{mon,mar}, a vacuum metric $g=\eta+\phi\,k\otimes k$, with $k$ null and geodesic, is paired with the abelian ``single copy'' $A^a=c^a\phi\,k$, which solves Maxwell's equations on flat space; the scalar $\phi$ is called the ``zeroth copy''. In the four-dimensional \emph{Weyl double copy} \cite{luna-weyl,godazgar-weyl}, the Weyl spinor of a vacuum spacetime of Petrov type D or N is written as $\Psi_{ABCD}=\phi_{(AB}\phi_{CD)}/S$, where $\phi_{AB}$ is a Maxwell field on that spacetime and $S$ is a scalar zeroth copy. Extensions to spacetimes with matter \cite{ems-sources,ems-em,amw} use several gauge fields and non-constant zeroth copies. Abelian precedents in which the physical source field also determines the Weyl curvature include the Reissner--Nordstr\"om and charged Born--Infeld self-doubling relations \cite{weyl-doubling}. Here we ask when the full Riemann tensor of an Einstein--Yang--Mills pp-wave equals the color-contracted square of its own Yang--Mills field strength with constant coefficient. For the solutions of Sec.~\ref{sec:pp}, this holds precisely for Coleman's profile with affine $V_0$.

\begin{theorem}\label{thm:dc}
Let $D\geq4$, let $g_{\mu\nu}$ be the pp-wave (\ref{ppwave}) with arbitrary profile $V(u,\vec y)$, and let $A^a_\mu=A^a\lambda_\mu$ with $\partial_vA^a=0$ and $\nabla^2A^a=0$. Let $\gamma_{ab}$ be positive definite. Then
\begin{equation}
R_{\mu\nu\rho\sigma}=\kappa\,\gamma_{ab}\,F^a_{\mu\nu}F^b_{\rho\sigma}
\label{exactDC}
\end{equation}
holds if and only if
\begin{equation}
A^a=p^a{}_i(u)y^i+c^a(u),\qquad V=-\frac{\kappa}{2}\gamma_{ab}A^aA^b+\beta_0(u)+\beta_i(u)y^i .
\end{equation}
In that case the Einstein equations $G_{\mu\nu}=\kappa T^{\rm YM}_{\mu\nu}$ hold.
\end{theorem}

\begin{proof}
The only independent components of the Riemann tensor of (\ref{ppwave}) are $R_{uiuj}=-\partial_i\partial_jV$. Those of the right-hand side are $\kappa\gamma_{ab}F^a_{ui}F^b_{uj}=\kappa\mathcal E_{ij}$, with $\mathcal E_{ij}\equiv\gamma_{ab}\partial_iA^a\partial_jA^b$, whose trace is the energy density $\mathcal E$; all other components of both sides vanish. So (\ref{exactDC}) is equivalent to $\partial_i\partial_jV=-\kappa\mathcal E_{ij}$. This has a solution $V$ if and only if $\partial_k\mathcal E_{ij}$ is symmetric in $(k,i)$. Write $\partial_k\mathcal E_{ij}=\mathcal S_{jki}+\mathcal S_{ikj}$ with $\mathcal S_{ikj}\equiv\gamma_{ab}\partial_iA^a\,\partial_k\partial_jA^b$, which is symmetric in its last two indices. Symmetry of $\partial_k\mathcal E_{ij}$ in $(k,i)$ is then equivalent to $\mathcal S_{ikj}=\mathcal S_{kij}$, so $\mathcal S$ is totally symmetric. Its trace on the last two indices is $\gamma_{ab}\partial_iA^a\nabla^2A^b=0$. By total symmetry the trace on the first two indices also vanishes:
\begin{equation*}
0=\mathcal S_{kkj}=\tfrac12\partial_j\big(\gamma_{ab}\nabla A^a\cdot\nabla A^b\big).
\end{equation*}
Hence $\mathcal E=\gamma_{ab}\nabla A^a\cdot\nabla A^b$ is independent of $\vec y$. Then $0=\nabla^2\mathcal E=2\gamma_{ab}\,\partial_i\partial_jA^a\,\partial_i\partial_jA^b$, and positivity of $\gamma_{ab}$ gives $\partial_i\partial_jA^a=0$. With $A^a$ affine, integrating $\partial_i\partial_jV=-\kappa E_{ij}$ gives the stated $V$. The converse is immediate. Finally, contracting (\ref{exactDC}) gives $R_{\nu\sigma}=\kappa\gamma_{ab}F^a{}_\nu{}^\rho F^b{}_{\sigma\rho}=\kappa T^{\rm YM}_{\nu\sigma}$, by Lemma~\ref{lem:null}(c), and $R=0$.
\end{proof}

Unlike Theorem~\ref{thm:unique}, Theorem~\ref{thm:dc} is a local statement: it needs no assumption about the behavior of the fields at infinity. A trace-free vacuum part $K_{ij}y^iy^j$ is excluded: a free gravitational wave superposed on the color wave spoils (\ref{exactDC}).

\begin{remark}[Weyl form]
The tensor $W_{\mu\nu\rho\sigma}\equiv\kappa\gamma_{ab}F^a_{\mu\nu}F^b_{\rho\sigma}$ has the symmetries of a curvature tensor. Its cyclic identity holds because $F^a\wedge F^b=0$ for null fields with a common $\lambda$. For any solution of the Einstein--Yang--Mills equations its Ricci-type contraction equals $R_{\mu\nu}$, by Einstein's equations and Lemma~\ref{lem:null}(c), and both scalar traces vanish. Hence, for such solutions, (\ref{exactDC}) is equivalent to the equality of Weyl parts, $C_{\mu\nu\rho\sigma}=W^{\rm Weyl}_{\mu\nu\rho\sigma}$. Theorem~\ref{thm:dc} can therefore be read as follows: \emph{among all Einstein--Yang--Mills pp-waves (\ref{guven}), whatever $V_0$, the Weyl tensor is the Weyl part of the color square of the field strength precisely for Coleman's profile with affine $V_0$.} In $D=4$ we use two-spinors as in \cite{penrose-rindler}. We write $F^a_{\mu\nu}\leftrightarrow\phi^a_{AB}\epsilon_{A'B'}+\bar\phi^a_{A'B'}\epsilon_{AB}$ and decompose the Riemann tensor as $R_{\mu\nu\rho\sigma}\leftrightarrow X_{ABCD}\epsilon_{A'B'}\epsilon_{C'D'}+\Phi_{ABC'D'}\epsilon_{A'B'}\epsilon_{CD}+\text{c.c.}$, where $X_{ABCD}=\Psi_{ABCD}+\frac{R}{24}(\epsilon_{AC}\epsilon_{BD}+\epsilon_{AD}\epsilon_{BC})$. Here $\Psi_{ABCD}$ is the Weyl spinor and $\Phi_{ABC'D'}$ the trace-free Ricci spinor. Comparing coefficients, (\ref{exactDC}) becomes
\begin{equation}
\Psi_{ABCD}=\kappa\,\gamma_{ab}\,\phi^a_{(AB}\phi^b_{CD)},\qquad
\Phi_{ABC'D'}=\kappa\,\gamma_{ab}\,\phi^a_{AB}\bar\phi^b_{C'D'},\qquad R=0 .
\end{equation}
For null fields, $\phi^a_{AB}=\phi^ao_Ao_B$, with complex amplitudes $\phi^a$ and a common spinor $o_A$ such that $\lambda_\mu\leftrightarrow o_A\bar o_{A'}$. Hence $\gamma_{ab}\phi^a_{AB}\phi^b_{CD}$ is already totally symmetric and has no trace part. Choose $m=(dx+i\,dy)/\sqrt2\leftrightarrow o_A\bar\iota_{A'}$ and $\epsilon_{AB}=o_A\iota_B-\iota_Ao_B$. Then $\phi^a=-\bar\varpi^a/\sqrt2$, and the two equations reproduce Eq.~(\ref{helicity}).
The first equation is a Weyl double copy with \emph{constant} zeroth copy $S=1/\kappa$, with colors contracted by $\gamma_{ab}$. The second is Einstein's equation. The Yang--Mills field here is the source of the spacetime, and it solves the Yang--Mills equations both on flat space and on the curved solution.
\end{remark}

\begin{remark}[Relation to other double copies]
The Kerr--Schild single copy of the G\"uven metric, $\tilde A^a_\mu=c^aV\lambda_\mu$, is abelian and obeys $\partial_\mu\tilde F^{\mu\nu}=-\kappa c^a\gamma_{bc}\nabla A^b\cdot\nabla A^c\,\lambda^\nu$. Its source is the energy density of the color wave, so it does not reproduce the gauge field that generated the metric. A double copy ``at the level of the Riemann tensor'' has recently been found for point sources in $2+1$ dimensions \cite{mrw}; there the gauge fields are again not the source of the curvature. Double-copy relations for scattering amplitudes on plane-wave backgrounds \cite{adamo-plane} and for vacuum type N Kundt spacetimes in arbitrary dimension \cite{opp2024} provide related constructions. Equation (\ref{exactDC}) is also not a double copy in the sense of Bern--Carrasco--Johansson \cite{bern,bcj2010}: color indices are contracted with $\gamma_{ab}$ rather than replaced by kinematic factors, and both sides belong to the same theory. Its content is that, for Coleman's profile, Einstein's equations determine the \emph{whole} Riemann tensor algebraically from $F^a_{\mu\nu}$, and not only its Ricci part.
\end{remark}

\section{Symmetry of the wave fronts and test particles}\label{sec:sym}

\subsection{Why Coleman's profile? A symmetry interpretation}\label{sec:symmetry}

Theorems~\ref{thm:unique} and~\ref{thm:dc} single out the same profile, so it is natural to ask whether a symmetry lies behind this. The metric and the field strength share the wave-front symmetries. For non-commuting amplitudes, the gauge potential cannot share all of them, even up to gauge; the reduced symmetry of Coleman waves was discussed by Trautman \cite{trautman-sym}.

\emph{Homogeneous wave fronts.} Call a vector $X$ \emph{tangent to the wave fronts} if $\lambda_\mu X^\mu=0$, i.e.\ if it is tangent to the null hypersurfaces $u=\text{const}$. On the pp-wave (\ref{ppwave}) the only non-zero Christoffel symbols are $\Gamma^v{}_{uu}=\partial_uV$, $\Gamma^v{}_{ui}=\partial_iV$ and $\Gamma^i{}_{uu}=-\partial_iV$. For any $v$-independent harmonic profile $A^a$ a short computation then gives $D_vF^a_{\mu\nu}=0$ and
\begin{equation}
D_kF^a_{uj}=-\partial_k\partial_jA^a ,
\end{equation}
with all other components of $D_kF^a_{\mu\nu}$ vanishing. Coleman's profile can therefore be characterized gauge-covariantly: \emph{$A^a$ is affine in $\vec y$ if and only if $D_XF_{\mu\nu}=0$ for every $X$ tangent to the wave fronts}. In words, the field strength is covariantly constant along each wave front. The gravitational counterpart is classical. Since $\nabla_kR_{uiuj}=-\partial_i\partial_j\partial_kV$ and $\nabla_vR_{\mu\nu\rho\sigma}=0$, the Riemann tensor is covariantly constant along the wave fronts if and only if $V$ is quadratic in $\vec y$, i.e.\ if and only if the pp-wave is a plane wave in the strict sense \cite{bpr}. The wave fronts of a plane wave are homogeneous. For the explicit symmetry formulas below, we work in Brinkmann coordinates with the affine terms in $V$ removed, and in a gauge with $c(u)=0$. Besides $\partial_v$, the metric admits the $2n$ Killing vectors \cite{bpr,zdgh}
\begin{equation}
\xi_\chi=\chi^i(u)\,\partial_i-\dot\chi_i(u)\,y^i\,\partial_v ,\qquad \ddot\chi_i=-R_{uiuj}\,\chi^j .
\label{heisenberg}
\end{equation}
The equation for $\chi$ is the geodesic deviation equation, so there is one Killing vector for each of the $2n$ independent transverse Jacobi fields. The Killing vectors close on $\partial_v$, $[\xi_\chi,\xi_{\tilde\chi}]=(\dot\chi\cdot\tilde\chi-\chi\cdot\dot{\tilde\chi})\,\partial_v$, which is the Heisenberg algebra, and their orbits are the wave fronts.

\begin{proposition}\label{prop:planewave}
The Einstein--Yang--Mills pp-wave (\ref{guven}) is a plane wave if and only if $A^a$ is Coleman's profile (\ref{coleman_A}), up to the removable term $c^a(u)$, and $V_0$ is quadratic in $\vec y$, up to affine terms.
\end{proposition}

\begin{proof}
If $\partial_i\partial_jV$ depends on $u$ only, so does its trace, $\nabla^2V=-\kappa\,\mathcal E$. As in the proof of Theorem~\ref{thm:dc}, $0=\nabla^2\mathcal E=2\gamma_{ab}\,\partial_i\partial_jA^a\,\partial_i\partial_jA^b$, and positivity of $\gamma_{ab}$ makes $A^a$ affine. Then $\gamma_{ab}A^aA^b$ is quadratic, and so is $V_0=V+\frac\kappa2\gamma_{ab}A^aA^b$. The converse is immediate.
\end{proof}

Like Theorem~\ref{thm:dc}, this proposition is local. Both theorems are thus statements about homogeneous wave fronts. Theorem~\ref{thm:unique} says that a bounded energy density must be uniform (this is where Liouville's theorem enters), and a uniform energy density is the matter part of the plane-wave condition. Theorem~\ref{thm:dc} is the full plane-wave condition with one extra requirement: there is no free gravitational wave, $K_{ij}=0$.

\emph{Symmetry of the potential.} The metric, the energy--momentum tensor and even the field strength are invariant under the Killing vectors (\ref{heisenberg}). Indeed $F^a=p^a{}_i\,dy^i\wedge du$ and $\pounds_{\xi_\chi}dy^i=\dot\chi^i\,du$, so $\pounds_{\xi_\chi}F^a=0$. The potential, however, need not be invariant as a one-form:
\begin{equation}
\pounds_{\xi_\chi}A=(\chi^ip_i)\,du .
\end{equation}
A gauge field is symmetric under an isometry $\xi$ if its Lie derivative is an infinitesimal gauge transformation, $\pounds_\xi A=D\Theta$ for some Lie-algebra-valued $\Theta$ \cite{forgacs-manton}. Here $D\Theta=d\Theta+[A,\Theta]$ may have only a $du$ component, so $\Theta=\Theta(u)$, and the condition becomes $\dot\Theta=\chi^ip_i$ together with $[p_i,\Theta]=0$ for all $i$. If the amplitudes lie in a fixed Cartan subalgebra (a maximal set of commuting generators), $\Theta=\int \chi^ip_i\,du$ solves this, and the wave is as homogeneous as a Maxwell plane wave. If the amplitudes do not commute, this cannot hold for all wave-front isometries, as the following gauge-invariant argument shows.

\begin{proposition}\label{prop:nosym}
If the Coleman--G\"uven wave is invariant up to gauge under all the wave-front isometries (\ref{heisenberg}), then $[p_i,p_j]=0$ for all $i,j$.
\end{proposition}

\begin{proof}
For Coleman's profile, the only non-vanishing components of $D_\alpha F_{\beta\gamma}$ are $D_uF_{ui}=-D_uF_{iu}=-\big(\dot p_i+[p_j,p_i]\,y^j\big)$. Hence
\begin{equation}
\mathcal J_{\alpha\beta\gamma\delta}\equiv\gamma_{ab}\,(D_\alpha F_{\beta\mu})^a\,(D_\gamma F_\delta{}^{\mu})^b=J\,\lambda_\alpha\lambda_\beta\lambda_\gamma\lambda_\delta ,\qquad
J=\sum_i\gamma_{ab}\big(\dot p_i+[p_j,p_i]y^j\big)^a\big(\dot p_i+[p_k,p_i]y^k\big)^b .
\end{equation}
The tensor $\mathcal J_{\alpha\beta\gamma\delta}$ is gauge invariant. If $\pounds_\xi A=D\Theta$, then $F$ and $DF$ transform by conjugation, so $\pounds_\xi\mathcal J_{\alpha\beta\gamma\delta}=0$. Since $\pounds_\xi\lambda=0$, this gives $\xi_\chi(J)=\chi^i\partial_iJ=0$. The value of $\chi$ at any given $u$ can be prescribed freely, so $J$ must be independent of $\vec y$. In particular its quadratic part, $\sum_i\gamma_{ab}[p_j,p_i]^a[p_k,p_i]^b\,y^jy^k$, must vanish. Positivity of $\gamma_{ab}$ then gives $[p_i,p_j]=0$.
\end{proof}

In $D=4$, for example, $J=\gamma_{ab}[p_1,p_2]^a[p_1,p_2]^b\,(x^2+y^2)$ plus terms at most linear in $x$ and $y$. A Coleman wave with non-commuting amplitudes therefore has less symmetry than its own field strength. The translated potential $A'=A+(\chi^ip_i)\,du$ has exactly the same field strength as $A$ at every point, because the added term is covariantly closed and its square vanishes (it is proportional to $du$). When the amplitudes do not commute, one can choose a wave-front translation for which $J[A']\neq J[A]$, proving that the two potentials are not gauge equivalent despite having the same field strength. This is a particularly simple instance of the ambiguity pointed out by Wu and Yang \cite{wu-yang} and by Deser and Wilczek \cite{deser-wilczek}: in a non-abelian theory, the field strength does not determine the potential up to gauge. It also gives a gauge-invariant meaning to the statement that Coleman's waves are genuinely non-abelian. For waves with non-commuting amplitudes, the geometry and the field strength are homogeneous along each wave front, but the potential is not, even up to gauge. Proposition~\ref{prop:nosym} gives a necessary condition for full wave-front symmetry; it does not identify pointwise commuting amplitudes with a potential valued in one fixed abelian subalgebra.

In the Poincar\'e patch of AdS (Sec.~\ref{sec:ads}), translations in $y^m$ are isometries, whereas translations in $z$ are not. AdS itself is homogeneous, with other isometries changing the radial coordinate. The radial dependence of the profile (\ref{adsA}) is determined by the Yang--Mills equation $\mathcal Q_2A^a=0$, rather than by translation symmetry alone; in $D=4$ this profile is linear in $z$ as well. The flat-space Liouville argument does not directly extend to this setting.

\subsection{Test particles: focusing and color memory}\label{sec:geodesics}

The symmetry discussed above has a direct physical consequence. Neutral test particles see homogeneous wave fronts, while colored test particles can detect the reduced symmetry of waves with non-commuting amplitudes.

\emph{Neutral particles.} Write the metric profile of the Coleman--G\"uven wave, with the removable affine terms dropped, as $V=-\frac12R_{uiuj}(u)\,y^iy^j$, with $R_{uiuj}=\kappa E_{ij}-2K_{ij}$ from (\ref{curvpp}). Since $\partial_v$ is a Killing vector, $\dot u$ is constant along every geodesic, so $u$ can be used as the parameter whenever $\dot u\neq0$. Denote $d/du$ by a prime. The transverse motion is then linear,
\begin{equation}
y^{i\,\prime\prime}=-R_{uiuj}(u)\,y^j ,
\label{geod}
\end{equation}
and $v$ follows from the normalization, $2v'=-\varsigma/\dot u^2-y'\cdot y'-2V$, with $\varsigma=1$ for massive and $\varsigma=0$ for massless particles. Equation (\ref{geod}) is the geodesic deviation equation. Its solutions are the same Jacobi fields $\chi^i(u)$ that define the Killing vectors (\ref{heisenberg}), and each $\xi_\chi$ gives a conserved quantity $\chi\cdot y'-\chi'\cdot y$. Geodesic motion is therefore completely integrable, which is the practical meaning of the homogeneity of the wave fronts.

Without a free gravitational wave ($K_{ij}=0$), Eq.~(\ref{geod}) reads $y^{i\,\prime\prime}=-\kappa E_{ij}(u)\,y^j$. Since $E_{ij}$ is positive semi-definite, the color wave acts as a converging lens whose strength along each eigenvector of $E_{ij}$ is the corresponding eigenvalue. A single Maxwell wave focuses only along its polarization direction; the conformally flat wave of Sec.~\ref{sec:pp} focuses equally in all directions, so a ring of particles stays circular as it shrinks.

For a sandwich wave of short duration, particles at rest at transverse positions $y_0$ before the wave leave it with velocities
\begin{equation}
\Delta y'=-\kappa\,\mathfrak E\,y_0 ,\qquad \mathfrak E_{ij}\equiv\int E_{ij}\,du ,
\end{equation}
to leading order in the duration. This is the velocity memory effect of plane waves \cite{zdgh}, which for color waves is always focusing. The inverse focal lengths $1/L_i$ are the eigenvalues of $\kappa\mathfrak E$, and their sum is fixed by the energy per unit transverse area, independently of the color structure:
\begin{equation}
\frac1{L_1}+\dots+\frac1{L_n}=\kappa\int\mathcal E\,du .
\end{equation}

\emph{Colored particles.} A test particle of mass $m$ carrying color charge $Q=Q^aT_a$ obeys Wong's equations \cite{wong},
\begin{equation}
m\,\frac{D\dot x^\mu}{d\tau}=\gamma_{ab}Q^aF^{b\,\mu}{}_\nu\,\dot x^\nu ,\qquad
\frac{dQ^a}{d\tau}+f^a{}_{bc}A^b_\mu\dot x^\mu Q^c=0 ,
\end{equation}
where the charge is normalized with $\gamma_{ab}$, consistently with $T^{\rm YM}_{\mu\nu}$ and with unit gauge coupling. For Coleman's profile the force has no $u$-component, so $\dot u$ is still constant, and the equations become
\begin{equation}
y^{i\,\prime\prime}=-R_{uiuj}\,y^j+\frac{1}{m\dot u}\,\gamma_{ab}Q^ap^b{}_i ,\qquad
Q'=-\big[y^jp_j,\,Q\big] .
\label{wong}
\end{equation}
Color-charge dynamics and radiation on Cartan-valued plane-wave backgrounds were studied in \cite{adamo-ilderton}. The norm $\gamma_{ab}Q^aQ^b$ is conserved, while the color charge precesses about $y^jp_j(u)$. The rate of precession depends on where the particle sits on the wave front. For a sandwich wave short enough that the particle hardly moves while the wave passes, the charge emerges rotated,
\begin{equation}
Q_{\rm out}=\mathcal U(y)\,Q_{\rm in}\,\mathcal U(y)^{-1} ,\qquad \mathcal U(y)=\mathcal P\exp\Big(-\int y^jp_j(u)\,du\Big) ,
\end{equation}
with later values of $u$ ordered to the left. Since $A_i=0$, transverse parallel transport is trivial, so it makes gauge-invariant sense to compare the charges of two particles at $y_1$ and $y_2$. If they agree before the wave, afterwards they differ by the relative rotation $\mathcal U(y_2)\,\mathcal U(y_1)^{-1}$. This is the color memory effect of Pate, Raclariu and Strominger \cite{prs}. Related path-ordered exponentials describe color memory in the shock waves of the color glass condensate \cite{bprsv}. Those fields share the null-potential ansatz, but their transverse profiles are sourced and generally not affine. The expression above is a formal solution for general amplitudes. For amplitudes in a fixed Cartan subalgebra, $\mathcal U(y_2)\mathcal U(y_1)^{-1}=\exp\big(-(y_2-y_1)^j\!\int p_j\,du\big)$ depends only on the separation. For non-commuting amplitudes the relative rotation generically depends on $y_1$ and $y_2$ separately. Through (\ref{wong}), the color force and transverse kick can likewise depend on the absolute transverse positions. The net color rotation can nevertheless vanish: if $p_i=f(u)T_i$ with fixed, possibly non-commuting $T_i$ and $\int f\,du=0$, the fixed-position expression above gives $\mathcal U(y)=1$. Colored matter can thus detect the inhomogeneity of the wave front that Proposition~\ref{prop:nosym} identified, even though neutral matter and the geometry are blind to it.

\section{Non-abelian waves in anti-de Sitter space}\label{sec:ads}

We now ask whether the construction survives in a curved background with a cosmological constant.

\subsection{Kerr--Schild--Kundt metrics}

Following \cite{gur1,gur6,gur-tek}, take
\begin{equation}
g_{\mu\nu}=\bar g_{\mu\nu}+2V\lambda_\mu\lambda_\nu ,
\label{ksk}
\end{equation}
where $\bar g_{\mu\nu}$ is the AdS metric of radius $\ell$, $\bar R_{\mu\nu}=-\frac{D-1}{\ell^2}\bar g_{\mu\nu}$, and $\lambda_\mu=\partial_\mu u$ and $V$ satisfy
\begin{equation}
\lambda^\mu\lambda_\mu=0,\qquad \nabla_\mu\lambda_\nu=\tfrac12(\xi_\mu\lambda_\nu+\xi_\nu\lambda_\mu),\qquad \xi_\mu\lambda^\mu=0,\qquad \lambda^\mu\partial_\mu V=0
\label{kskprop}
\end{equation}
for some covector $\xi_\mu$. These relations hold for both the AdS and the full connection. They imply that the integral curves of $\lambda^\mu$ are null geodesics with vanishing expansion, shear and twist. In other words, a bundle of light rays along $\lambda$ neither spreads, nor distorts, nor rotates; spacetimes with such a null congruence are called Kundt spacetimes. For geodesic Kerr--Schild vectors, the mixed Einstein tensor is already linear in the profile \cite{Gurses_Gursey,gergely}; see also \cite{Dereli} for generalized Kerr--Schild transformations. The full KSK conditions above restrict its deviation from the AdS Einstein tensor to the single $\lambda^\mu\lambda_\nu$ structure displayed below. Metrics (\ref{ksk})--(\ref{kskprop}) are called Kerr--Schild--Kundt (KSK) metrics. The Einstein tensor is \cite{gur1}
\begin{equation}
G_{\mu\nu}=\frac{(D-1)(D-2)}{2\ell^2}g_{\mu\nu}-\lambda_\mu\lambda_\nu\mathcal Q_1V,\qquad
\mathcal Q_1\equiv\bar\Box+2\xi^\mu\partial_\mu+\frac12\xi^\mu\xi_\mu-\frac{2(D-2)}{\ell^2}.
\end{equation}

\subsection{Reduced field equations}

For the null gauge field (\ref{ansatz}), Lemma~\ref{lem:null} applies. In (\ref{divF}), $\nabla_\mu\lambda_\nu+\nabla_\nu\lambda_\mu=\xi_\mu\lambda_\nu+\xi_\nu\lambda_\mu$ and $\nabla\cdot\lambda=\xi\cdot\lambda=0$. Moreover $\Box A^a=\bar\Box A^a$, because the Kerr--Schild term leaves $\sqrt{-g}$ unchanged and its correction to the inverse metric, $-2V\lambda^\mu\lambda^\nu$, annihilates $\partial A^a$. Hence $\nabla^\mu F^a_{\mu\nu}=(\mathcal Q_2A^a)\lambda_\nu$ with $\mathcal Q_2\equiv\bar\Box+\xi^\mu\partial_\mu$. For AdS plane waves, both reduced equations were given in \cite{cggp} for arbitrary $D\geq4$; the general KSK formulation used below was developed in \cite{gur-tek}.

\begin{proposition}\label{prop:ksk}
For the ansatz (\ref{ansatz}) on the KSK metric (\ref{ksk}), the equations $G_{\mu\nu}+\Lambda g_{\mu\nu}=\kappa T^{\rm YM}_{\mu\nu}$ and $(D_\mu F^{\mu\nu})^a=0$ reduce to
\begin{equation}
\mathcal Q_2A^a=0,\qquad
\mathcal Q_1V=-\kappa\,\bar g^{\alpha\beta}\gamma_{ab}\,\partial_\alpha A^a\partial_\beta A^b,\qquad
\Lambda=-\frac{(D-1)(D-2)}{2\ell^2}.
\label{kskred}
\end{equation}
\end{proposition}

\subsection{Exact AdS plane waves}

In Poincar\'e coordinates,
\begin{equation}
ds^2=\frac{\ell^2}{z^2}\Big(2\,du\,dv+\sum_{m=1}^{D-3}(dy^m)^2+dz^2\Big)+2V(u,y^m,z)\,du^2,\qquad
\lambda_\mu dx^\mu=du,\qquad \xi_\mu dx^\mu=\frac2z\,dz ,
\label{poincare}
\end{equation}
with $\xi^\mu\xi_\mu=4/\ell^2$. The conformal boundary is at $z=0$ and the Poincar\'e horizon at $z\to\infty$. On $v$-independent functions, with $\tilde\partial^2\equiv\sum_m\partial_{y^m}^2+\partial_z^2$,
\begin{equation}
\mathcal Q_2=\frac{z^2}{\ell^2}\Big(\tilde\partial^2-\frac{D-4}{z}\partial_z\Big),\qquad
\mathcal Q_1=\frac{z^2}{\ell^2}\Big(\tilde\partial^2-\frac{D-6}{z}\partial_z\Big)-\frac{2(D-3)}{\ell^2}.
\label{Qops}
\end{equation}
The vacuum equation $\mathcal Q_1V_0=0$ describes the gravitational AdS waves of Siklos \cite{siklos,podolsky1998}: writing $V_0=\ell^2\psi/z^2$, one has $\mathcal Q_1V_0=(\ell^2/z^2)\,\bar\Box\psi$. Particular solutions are easy to find because powers of $z$ are eigenfunctions:
\begin{equation}
\mathcal Q_1z^r=\Omega_r\,z^r,\qquad \Omega_r=\frac{(r-D+3)(r+2)}{\ell^2},
\label{Kalpha}
\end{equation}
for any constant $r$.

\begin{proposition}\label{prop:ads}
Let $D\geq4$. The gauge profile
\begin{equation}
A^a=p^a{}_m(u)\,y^m+q^a(u)\,z^{D-3}
\label{adsA}
\end{equation}
solves $\mathcal Q_2A^a=0$. Define
\begin{equation}
E_y(u)\equiv\delta^{mm'}\gamma_{ab}p^a{}_m(u)p^b{}_{m'}(u),\qquad E_z(u)\equiv\gamma_{ab}q^a(u)q^b(u),
\end{equation}
where $m,m'=1,\ldots,D-3$ run over the coordinates $y^m$ only. The metric profile
\begin{equation}
V=\frac{\kappa\,z^2}{4(D-5)}\,E_y(u)-\frac{\kappa\,(D-3)\,z^{2(D-3)}}{2(D-2)}\,E_z(u)+V_0,\qquad \mathcal Q_1V_0=0,
\label{adsV}
\end{equation}
solves (\ref{kskred}) for $D\neq5$. For $D=5$ the first term is replaced by $-\frac{\kappa}{4}E_y(u)\,z^2\ln\frac z\ell$.
\end{proposition}

\begin{proof}
Both terms of (\ref{adsA}) are annihilated by $\mathcal Q_2$, the second because $\partial_z^2z^{D-3}=\frac{D-4}{z}\partial_zz^{D-3}$. Since $\bar g^{\alpha\beta}$ is diagonal in $(y^m,z)$, the source is
\begin{equation*}
\bar g^{\alpha\beta}\gamma_{ab}\partial_\alpha A^a\partial_\beta A^b=\frac{z^2}{\ell^2}\,E_y(u)+\frac{(D-3)^2\,z^{2(D-3)}}{\ell^2}\,E_z(u) .
\end{equation*}
Dividing by $\Omega_2=-4(D-5)/\ell^2$ and by $\Omega_{2(D-3)}=2(D-2)(D-3)/\ell^2$ gives (\ref{adsV}). For $D=5$, $\Omega_2=0$: $z^2$ solves the vacuum equation, and $\mathcal Q_1(z^2\ln\frac z\ell)=\frac4{\ell^2}z^2$ supplies the particular solution.
\end{proof}

The field strength has components $F^a_{um}=-p^a{}_m$ and $F^a_{uz}=-(D-3)z^{D-4}q^a$, with
\begin{equation}
[F_{um},F_{um'}]^a=f^a{}_{bc}p^b{}_mp^c{}_{m'},\qquad [F_{um},F_{uz}]^a=(D-3)\,z^{D-4}f^a{}_{bc}p^b{}_mq^c .
\end{equation}
The solution is therefore non-abelian in every $D\geq4$ as soon as some $p_m$ fails to commute with $q$ (or, for $D\geq5$, with another $p_{m'}$). In four dimensions, with $y\equiv y^1$ and $p^a\equiv p^a{}_1$, it reads
\begin{equation}
A^a=p^a(u)\,y+q^a(u)\,z,\qquad V=-\frac{\kappa}{4}\,\gamma_{ab}\big(p^ap^b+q^aq^b\big)z^2+V_0 .
\label{ads4}
\end{equation}
In $D=4$, $\mathcal Q_2=\frac{z^2}{\ell^2}\tilde\partial^2$, so every solution of the flat Laplace equation in $(y,z)$ lifts to AdS$_4$. This reflects the conformal invariance of four-dimensional Yang--Mills theory together with the conformal flatness of AdS$_4$.

\subsection{Two differences from flat space}

\emph{Energy density.} Set $V_0=0$ in (\ref{ads4}) and write $\mathcal E_0\equiv E_y+E_z=\gamma_{ab}(p^ap^b+q^aq^b)$. Static observers in Poincar\'e coordinates move along $\partial_t$, where $t=(v-u)/\sqrt2$ and $w=(u+v)/\sqrt2$, so that $2\,du\,dv=-dt^2+dw^2$. They measure
\begin{equation}
\mathcal E=\frac{z^4\,\mathcal E_0}{2\ell^2\big(\ell^2+\kappa\,\mathcal E_0z^4/4\big)} .
\end{equation}
Neglecting the gravity of the wave, $\mathcal E\simeq z^4\mathcal E_0/(2\ell^4)$ would grow without bound toward the Poincar\'e horizon. The exact energy density is instead bounded by $2/(\kappa\ell^2)$, a bound independent of the amplitude of the wave. The reason is that the wave's own gravitational field ($V<0$, growing like $z^2$) stretches the proper time of the static observers, and this reduces their Doppler factor $(\lambda\cdot U)^2$. The bound therefore refers to these observers: boosted observers measure any value, as in flat space. The boundedness is not specific to $D=4$. For $V_0=0$, $E_y=0$ and $E_z>0$, the static-observer energy density increases monotonically with $z$, at fixed $u$, toward $(D-3)(D-2)/(\kappa\ell^2)$ in every $D\geq4$. For $E_y>0$ and $D\geq5$, the factor $\ell^2-z^2V$ can vanish at finite $z$; $\partial_t$ then becomes null, and the density measured by normalized static observers diverges on approach from the timelike region. For $D\geq6$ and $E_z=0$, such a zero always occurs. Theorem~\ref{thm:unique} has no direct analog here, because Liouville's theorem does not hold on the half-space $z>0$.

\emph{Failure of the flat-space curvature identity.} Equation~(\ref{exactDC}) cannot hold for these AdS waves in any dimension: its double trace would require $R=\kappa\gamma_{ab}F^a_{\mu\nu}F^{b\,\mu\nu}=0$, whereas $R=-D(D-1)/\ell^2$. In $D=4$ we can say more. With $V_0=0$, the independent components of the Weyl tensor of (\ref{ads4}), and of the Weyl part of $W_{\mu\nu\rho\sigma}=\kappa\gamma_{ab}F^a_{\mu\nu}F^b_{\rho\sigma}$, are
\begin{align}
&C_{uyuy}=-C_{uzuz}=-\frac{3\kappa}{2}\gamma_{ab}\big(p^ap^b+q^aq^b\big),&& C_{uyuz}=0,\nonumber\\
&W^{\rm Weyl}_{uyuy}=-W^{\rm Weyl}_{uzuz}=\frac\kappa2\gamma_{ab}\big(p^ap^b-q^aq^b\big),&& W^{\rm Weyl}_{uyuz}=\kappa\gamma_{ab}p^aq^b .
\end{align}
In AdS$_4$ the Weyl tensor tracks the total energy of the wave, while $W^{\rm Weyl}$ tracks its anisotropy. In particular, for orthonormal color amplitudes ($\gamma_{ab}p^ap^b=\gamma_{ab}q^aq^b$, $\gamma_{ab}p^aq^b=0$, which for $su(2)$ is a maximally non-abelian choice) $W^{\rm Weyl}=0$ but $C\neq0$. There is thus no relation $C=\varphi\,W^{\rm Weyl}$ with $\varphi$ independent of the color amplitudes, and the analog of Theorem~\ref{thm:dc} fails.

\section{Waves on \texorpdfstring{$\mathbb R^{1,1}\times S^{D-2}$}{R(1,1) x S(D-2)}}\label{sec:sphere}

Replace the flat transverse space of Sec.~\ref{sec:pp} by an $n$-dimensional Riemannian space $(N,h)$ of constant curvature $k$:
\begin{equation}
ds^2=2\,du\,dv+2V(u,y)\,du^2+h_{ij}(y)\,dy^idy^j ,\qquad \bar R_{ijkl}=k(h_{ik}h_{jl}-h_{il}h_{jk}),\qquad n=D-2 .
\label{ppcc}
\end{equation}
In $D=4$ such metrics were found as non-Einstein solutions of quadratic gravity \cite{heefer}, with spherical and with hyperbolic wave fronts, and they are almost universal \cite{gur-sis-tek}. We restrict to constant curvature: for an Einstein space that is not of constant curvature ($n\geq4$), the Weyl tensor of $h$ spoils the structure of the generic field equations used in Sec.~\ref{sec:generic}. The vector $\lambda=du$ is still null and covariantly constant. With $h_{\mu\nu}$ denoting $h_{ij}$ extended by zeros, the Einstein tensor is
\begin{equation}
G_{\mu\nu}=(n-1)k\,h_{\mu\nu}-\frac{n(n-1)k}{2}\,g_{\mu\nu}-(\Delta_hV)\,\lambda_\mu\lambda_\nu ,
\label{Gcc}
\end{equation}
where $\Delta_h$ is the Laplacian of $h$.

For $k\neq0$ the term proportional to $h_{\mu\nu}$ must be supported by matter. Following \cite{ortag01} we use the Maxwell field $\mathcal A=\varepsilon\,v\,du$, with constant $\varepsilon$. Its field strength $\mathcal F=\varepsilon\,dv\wedge du$ has the single constant contravariant component $\mathcal F^{uv}$, and $\sqrt{-g}=\sqrt h$, so Maxwell's equations hold. Its energy--momentum tensor is independent of $V$:
\begin{equation}
T^{(\mathcal F)}_{\mu\nu}=\varepsilon^2\Big(h_{\mu\nu}-\frac12g_{\mu\nu}\Big).
\end{equation}
In $D=4$ the background ($V=0$) is the Pleba\'nski--Hacyan direct product \cite{plebanski-hacyan} of two-dimensional Minkowski space and a sphere, and pp-waves on it are known \cite{ortag01}. For the null Yang--Mills field (\ref{ansatz}), Lemma~\ref{lem:null} gives $\Box A^a=\Delta_hA^a$ and $T^{\rm YM}_{\mu\nu}=\gamma_{ab}h^{ij}\partial_iA^a\partial_jA^b\,\lambda_\mu\lambda_\nu$.

\begin{proposition}\label{prop:cc}
The Einstein--Maxwell--Yang--Mills equations $G_{\mu\nu}+\Lambda g_{\mu\nu}=\kappa T^{(\mathcal F)}_{\mu\nu}+\kappa T^{\rm YM}_{\mu\nu}$, $\nabla_\mu\mathcal F^{\mu\nu}=0$, $(D_\mu F^{\mu\nu})^a=0$ reduce to
\begin{align}
&(n-1)k=\kappa\varepsilon^2,\qquad \Lambda=\frac{n(n-1)k}{2}-\frac{\kappa\varepsilon^2}{2}=\frac{(n-1)^2}{2}\,k,\label{eq1}\\
&\Delta_hA^a=0,\qquad \Delta_hV=-\kappa\,\gamma_{ab}h^{ij}\partial_iA^a\partial_jA^b .\label{eq2}
\end{align}
Hence $k\geq0$, with $k>0$ (and then $\Lambda>0$) if and only if $\varepsilon\neq0$. The general solution is $V=V_0-\frac{\kappa}{2}\gamma_{ab}A^aA^b$ with $\Delta_hV_0=0$. For $k=0$ one needs $\varepsilon=0$, and Sec.~\ref{sec:pp} is recovered.
\end{proposition}

For a sphere of radius $a_0=k^{-1/2}$ one has $\kappa\varepsilon^2=(D-3)/a_0^2$ and $\Lambda=(D-3)^2/(2a_0^2)$.

\begin{proposition}[No regular waves]\label{prop:noreg}
Let $N$ be complete and connected, with $n\geq2$ and $k>0$, and let $A^a(u,\cdot)$ and $V_0(u,\cdot)$ be $C^2$ on $N$. Then $F^a_{\mu\nu}=0$, and $V_0$ depends on $u$ only and can be removed by a coordinate transformation.
\end{proposition}

\begin{proof}
By Myers' theorem \cite{myers} $N$ is compact; it is the round sphere $S^{n}$ or a quotient of it. Multiplying $\Delta_hA^a=0$ by $\gamma_{ab}A^b$ and integrating by parts gives $\int_N\sqrt h\,\gamma_{ab}h^{ij}\partial_iA^a\partial_jA^b=0$. Since $\gamma_{ab}$ is positive definite, $\partial_iA^a=0$. Then $A^a=A^a(u)$ and, by Lemma~\ref{lem:null}(a), $F^a=0$. The same argument shows that $V_0=V_0(u)$, which the shift $v\to v-\int V_0\,du$ removes.
\end{proof}

Hence every non-trivial wave on this background is singular somewhere on the sphere. An example on $S^2$: with $s=\ln\tan\frac\theta2$, the metric is $h=a_0^2\sin^2\theta\,(ds^2+d\phi^2)$, so away from the poles the harmonic equation reduces to that on the flat cylinder with coordinates $(s,\phi)$. The profile
\begin{equation}
A^a=p^a(u)\,s+q^a(u)\,e^s\cos\phi
\end{equation}
is single-valued and harmonic away from the poles, and is non-abelian where $[p,q]\neq0$. Together with $V=-\frac\kappa2\gamma_{ab}A^aA^b$ it solves the full system, with energy density
\begin{equation}
\gamma_{ab}h^{ij}\partial_iA^a\partial_jA^b=\frac{\gamma_{ab}p^ap^b+2e^s\cos\phi\,\gamma_{ab}p^aq^b+e^{2s}\gamma_{ab}q^aq^b}{a_0^2\sin^2\theta}.
\end{equation}
The $p$ term diverges at both poles, the $q$ term at $\theta=\pi$.

\section{Generic gravity and gauge theories}\label{sec:generic}

Consider any metric theory of gravity whose Lagrangian is built from the metric, the Riemann tensor and its covariant derivatives. Write its field equations as
\begin{equation}
G_{\mu\nu}+\Lambda g_{\mu\nu}+H_{\mu\nu}=\kappa T_{\mu\nu},
\label{eq:generic-gravity}
\end{equation}
where $H_{\mu\nu}$ collects the contributions of the higher-curvature terms. A metric is called \emph{universal} if every symmetric, divergence-free rank-two tensor constructed locally from the metric, the Riemann tensor and its covariant derivatives is proportional to the metric, with a constant coefficient \cite{coley,hpp2014}. Consequently, the source-free field equations of any such metric theory reduce to a single algebraic condition. For a universal metric, the higher-curvature correction takes the form $H_{\mu\nu}=e_0 g_{\mu\nu}$, so it only shifts the cosmological term, $\Lambda\to\Lambda+e_0$. \emph{Almost universality} relaxes this condition: for every such theory, the field equations reduce to a finite set of algebraic conditions and a single linear scalar equation for a profile function \cite{gur-sis-tek}. For the metrics considered here, $H_{\mu\nu}$ reduces to a combination of a few fixed tensor structures, with theory-dependent constants and a linear differential operator acting on the profile $V$. We give these structures explicitly below.

\subsection{pp-waves and AdS waves}\label{sec:generic-ppksk}

For pp-wave and KSK metrics \cite{ghst, gur4,gur6},
\begin{equation}
H_{\mu\nu}=e_0\,g_{\mu\nu}+\lambda_\mu\lambda_\nu\sum_{i=1}^MC_i\,\mathcal Q_1^{\,i}V ,
\end{equation}
with constants $e_0$ and $C_i$ that depend on the theory and on $\ell$; for pp-waves $\mathcal Q_1\to\nabla^2$ and $e_0=0$. The sum starts at $i=1$. For pp-waves, the coordinate change $v^{\prime}=v-\int c(u)\,du$ gives $V^{\prime}=V+c(u)$, and $\nabla^2c(u)=0$. For AdS plane waves in the Poincar\'e coordinates (\ref{poincare}), the same change gives $V^{\prime}=V+\ell^2c(u)/z^2$, with $\mathcal Q_1[\ell^2c(u)/z^2]=0$ by (\ref{Kalpha}). Covariance therefore excludes an additional $C_0V$ term outside the powers of $\mathcal Q_1$ in the $\lambda\lambda$ part. In the general KSK tensor reduction, the factor of $\mathcal Q_1$ comes from the traceless Ricci tensor, $R_{\mu\nu}-\tfrac{R}{D}g_{\mu\nu}=-\lambda_\mu\lambda_\nu\mathcal Q_1V$. To write the resulting linear operator acting on $V$ compactly, define the polynomial
\begin{equation}
P(x)\equiv x-\sum_{i=1}^MC_i\,x^i .
\end{equation}
Thus $P(x)=x$ in Einstein's theory, $P(0)=0$ always, and $P'(0)=1-C_1$. For the null gauge ansatz (\ref{ansatz}) on the pp-wave metric (\ref{ppwave}) and the AdS KSK metric (\ref{ksk}), the Yang--Mills equations reduce to $\nabla^2A^a=0$ and $\mathcal Q_2A^a=0$, respectively, as established in (\ref{ppred}) and (\ref{kskred}). These equations, and the Yang--Mills source on the right-hand side of (\ref{eq:generic-gravity}), are unchanged, because the higher-curvature corrections modify only the gravitational sector. Substituting $H_{\mu\nu}$ into (\ref{eq:generic-gravity}) and equating the coefficients of the independent tensors $g_{\mu\nu}$ and $\lambda_\mu\lambda_\nu$ gives, respectively, the background condition and the equation for $V$. For AdS waves, the former changes the Einstein relation $\Lambda=-(D-1)(D-2)/(2\ell^2)$ to $\Lambda+e_0=-(D-1)(D-2)/(2\ell^2)$. Since $e_0$ depends on the gravitational couplings and on $\ell$, this is an algebraic condition determining the allowed AdS radii for a given theory and cosmological constant. For pp-waves, $e_0=0$ and $\Lambda=0$. The profile equation replaces $\nabla^2$ by $P(\nabla^2)$ for pp-waves and $\mathcal Q_1$ by $P(\mathcal Q_1)$ for AdS waves. Thus the reduced system becomes
\begin{align}
\text{pp-waves:}&\quad P(\nabla^2)V=-\kappa\,\gamma_{ab}\nabla A^a\!\cdot\!\nabla A^b,\qquad \nabla^2A^a=0,\qquad \Lambda=0;\\
\text{AdS waves:}&\quad P(\mathcal Q_1)V=-\kappa\,\bar g^{\alpha\beta}\gamma_{ab}\partial_\alpha A^a\partial_\beta A^b,\qquad
\mathcal Q_2A^a=0,\qquad \Lambda=-e_0-\frac{(D-1)(D-2)}{2\ell^2}.
\end{align}

For Coleman's profile (\ref{coleman_A}) and its AdS generalization (\ref{adsA}), the solutions of these equations take the following explicit forms.

\begin{proposition}\label{prop:generic}
(a) For Coleman's profile (\ref{coleman_A}) in a pp-wave,
\begin{equation}
V=-\frac{\kappa}{2P'(0)}\,\gamma_{ab}A^aA^b+V_0,\qquad P(\nabla^2)V_0=0,
\end{equation}
provided $P'(0)\neq0$. With $V_0$ affine, $R_{\mu\nu\rho\sigma}=\frac{\kappa}{P'(0)}\gamma_{ab}F^a_{\mu\nu}F^b_{\rho\sigma}$: the curvature double copy of Theorem~\ref{thm:dc} persists with an effective coupling.\\
(b) For the AdS profile (\ref{adsA}),
\begin{equation}
V=-\frac{\kappa\,z^2}{P(\Omega_2)\,\ell^2}\,E_y(u)-\frac{\kappa\,(D-3)^2\,z^{2(D-3)}}{P(\Omega_{2(D-3)})\,\ell^2}\,E_z(u)+V_0,\qquad P(\mathcal Q_1)V_0=0,
\end{equation}
provided the denominators do not vanish. For $D=5$, where $\Omega_2=0$, the first term is replaced by $-\frac{\kappa}{4P'(0)}E_y(u)z^2\ln\frac z\ell$. Here, $\Omega$ is defined in (\ref{Kalpha}). 
\end{proposition}

\begin{proof}
(a) $\nabla^2(\gamma_{ab}A^aA^b)=2E(u)$ is independent of $\vec y$, so all higher powers of $\nabla^2$ annihilate $\gamma_{ab}A^aA^b$ and $P(\nabla^2)$ acts on it as $P'(0)\nabla^2$. The curvature statement follows as in Theorem~\ref{thm:dc}. (b) $P(\mathcal Q_1)z^r=P(\Omega_r)z^r$ by (\ref{Kalpha}). For $D=5$, $\mathcal Q_1(z^2\ln\frac z\ell)=\frac4{\ell^2}z^2$ and $\mathcal Q_1z^2=0$, so $P(\mathcal Q_1)$ acts on $z^2\ln\frac z\ell$ as $P'(0)\mathcal Q_1$.
\end{proof}

For $P(x)=x$ these formulas reduce to (\ref{guven}) and (\ref{adsV}). A vanishing denominator signals a resonance: the corresponding power of $z$ then solves the vacuum equation, and the particular solution acquires a logarithm, exactly as in $D=5$. In $D=5$ this resonance is present in every theory. For Lovelock gravity, the same structure, a single effective coupling multiplying the Einstein operator, was found earlier by Gleiser and Dotti \cite{gleiser-dotti}.

For pp-waves with non-affine harmonic gauge profiles, $\nabla^4(\gamma_{ab}A^aA^b)=4\gamma_{ab}\partial_i\partial_jA^a\partial_i\partial_jA^b$ is not identically zero. Higher-curvature coefficients beyond $C_1$ can therefore affect the particular solution, but which coefficients contribute depends on the profile; non-affineness alone does not imply that all $C_i$ enter.

\subsection{pp-waves over constant-curvature spaces}

For the metrics (\ref{ppcc}) \cite{gur-sis-tek},
\begin{equation}
H_{\mu\nu}=e_0\,g_{\mu\nu}+e_1\,h_{\mu\nu}+\lambda_\mu\lambda_\nu\sum_{i=1}^MC_i\,\Delta_h^{\,i}V ,
\end{equation}
with constants $e_0$, $e_1$, $C_i$ depending on the theory and on $k$. As before, no term proportional to $V$ itself appears in the $\lambda\lambda$ part. Equations (\ref{eq1})--(\ref{eq2}) become
\begin{align}
&(n-1)k+e_1=\kappa\varepsilon^2,\qquad \Lambda+e_0=\frac{n(n-1)k}{2}-\frac{\kappa\varepsilon^2}{2},\\
&\Delta_hA^a=0,\qquad P(\Delta_h)V=-\kappa\,\gamma_{ab}h^{ij}\partial_iA^a\partial_jA^b .
\end{align}
The Yang--Mills conclusion of Proposition~\ref{prop:noreg} remains unchanged: for $k>0$ on a complete connected transverse space of dimension $n\geq2$, every globally regular harmonic profile $A^a$ is independent of the transverse coordinates, and hence $F^a_{\mu\nu}=0$. Consequently, a non-trivial Yang--Mills wave in our ansatz must be singular somewhere on this compact transverse space. Higher-curvature corrections to the gravitational equations cannot remove this obstruction: it follows from the unchanged Yang--Mills equation and is independent of the polynomial $P$.

The constant $e_1$, however, could allow $k<0$: a non-compact hyperbolic transverse space, on which bounded-energy harmonic gauge profiles exist in abundance. We now show that coupled waves require further conditions.

\subsection{Can a hyperbolic transverse space help?}\label{sec:hyp}

Let the background $\mathbb R^{1,1}\times N^n_k$ be supported by any matter that respects the product structure, $T_{\mu\nu}=t_h\,h_{\mu\nu}+t_g\,g_{\mu\nu}$ with constants $t_h,t_g$, plus the Yang--Mills wave. The \emph{null energy condition}, $T_{\mu\nu}w^\mu w^\nu\geq0$ for all null $w$, is equivalent to $t_h\geq0$: take $w=\partial_t+\hat e$ with $\hat e$ a unit vector tangent to $N$. The Maxwell field above has $t_h=\varepsilon^2$. Consider quadratic and Gauss--Bonnet gravity,
\begin{equation}
\frac1\kappa\big(G_{\mu\nu}+\Lambda g_{\mu\nu}\big)+\alpha_1\mathcal H^{(1)}_{\mu\nu}+\alpha_2\mathcal H^{(2)}_{\mu\nu}+\alpha_3\mathcal H^{\rm GB}_{\mu\nu}=T_{\mu\nu},
\end{equation}
so that $H_{\mu\nu}=\kappa(\alpha_1\mathcal H^{(1)}_{\mu\nu}+\alpha_2\mathcal H^{(2)}_{\mu\nu}+\alpha_3\mathcal H^{\rm GB}_{\mu\nu})$. Here $\mathcal H^{(1)}$, $\mathcal H^{(2)}$ and $\mathcal H^{\rm GB}$ are the variational derivatives of $R^2$, $R_{\mu\nu}R^{\mu\nu}$ and the Gauss--Bonnet density $\mathcal G=R^2-4R_{\mu\nu}R^{\mu\nu}+R_{\mu\nu\rho\sigma}R^{\mu\nu\rho\sigma}$:
\begin{align*}
\mathcal H^{(1)}_{\mu\nu}&=2RR_{\mu\nu}-\tfrac12g_{\mu\nu}R^2+2(g_{\mu\nu}\Box-\nabla_\mu\nabla_\nu)R,\\
\mathcal H^{(2)}_{\mu\nu}&=\Box R_{\mu\nu}+\tfrac12g_{\mu\nu}\Box R-\nabla_\mu\nabla_\nu R+2R_{\mu\sigma\nu\rho}R^{\sigma\rho}-\tfrac12g_{\mu\nu}R_{\sigma\rho}R^{\sigma\rho},\\
\mathcal H^{\rm GB}_{\mu\nu}&=2\big(RR_{\mu\nu}-2R_{\mu\alpha}R^\alpha{}_\nu-2R_{\mu\alpha\nu\beta}R^{\alpha\beta}+R_\mu{}^{\alpha\beta\gamma}R_{\nu\alpha\beta\gamma}\big)-\tfrac12g_{\mu\nu}\mathcal G .
\end{align*}
The $h_{\mu\nu}$ and $\lambda_\mu\lambda_\nu$ parts of the field equations are
\begin{align}
&\frac{(n-1)k}{\kappa}\,X_h=t_h ,\label{hyp-h}\\
&-\frac{X_V}{\kappa}\,\Delta_hV-\alpha_2\,\Delta_h^2V=\gamma_{ab}\,h^{ij}\partial_iA^a\partial_jA^b ,\qquad \Delta_hA^a=0,\label{hyp-V}
\end{align}
with
\begin{equation}
X_h=Y+2\kappa\alpha_3(n-2)(n-3)k,\qquad
X_V=Y+2\kappa\alpha_3(n-1)(n-2)k,\qquad
Y\equiv1+2\kappa(n-1)k\,(n\alpha_1+\alpha_2),
\label{XhXV}
\end{equation}
while the $g_{\mu\nu}$ part fixes $\Lambda$. In the notation of Sec.~\ref{sec:generic-ppksk}, $P(x)=X_Vx+\kappa\alpha_2x^2$. For $\alpha_2=0$, $Y=f'(\bar R)$ is the familiar factor of $f(R)$ gravity. In quadratic theories ($\alpha_3=0$) $X_h=X_V$; only the Gauss--Bonnet term distinguishes the effective coupling of the background from that of the waves, $X_V-X_h=4\kappa\alpha_3(n-2)k$.

We say the waves \emph{anti-gravitate} if $R_{uu}$ and $T^{\rm YM}_{uu}$ have opposite signs, i.e.\ if their energy defocuses, rather than focuses, light rays crossing them.

\begin{theorem}\label{thm:hyp}
Let the matter supporting $\mathbb R^{1,1}\times N^n_k$ satisfy the null energy condition, $t_h\geq0$. Then:
\begin{enumerate}
\item[(a)] In Einstein gravity with $t_h>0$, $k>0$, so $N$ is compact and every non-trivial wave is singular. This holds for any such matter, not only for the Maxwell field.
\item[(b)] Let $t_h>0$ and $\alpha_3\geq0$ (in particular in all quadratic theories, in $f(R)$ gravity, and in Einstein--Gauss--Bonnet gravity with the sign of string theory). Then $k<0$ implies $X_V\leq X_h<0$: the effective coupling of the waves, $\kappa/X_V$, is negative. For $\alpha_2=0$ this means $R_{uu}=(\kappa/X_V)\,T^{\rm YM}_{uu}<0$: the waves anti-gravitate.
\item[(c)] Let $t_h>0$. A hyperbolic transverse space with $X_V>0$ requires $\alpha_3<0$, $D\geq5$ and $Y<0$; for $\alpha_2=0$ the last condition means $f'(\bar R)<0$.
\item[(d)] If $t_h=0$ (no supporting matter) and $k\neq0$, then $X_h=0$, i.e.\ $|k|=1/\big(2\kappa\,|(n-1)(n\alpha_1+\alpha_2)+(n-2)(n-3)\alpha_3|\big)$.
\end{enumerate}
\end{theorem}

\begin{proof}
(a) $X_h=1$, so $(n-1)k=\kappa t_h>0$, and Proposition~\ref{prop:noreg} applies. (b) By (\ref{hyp-h}), $k<0$ and $t_h>0$ force $X_h<0$, and $X_V-X_h=4\kappa\alpha_3(n-2)k\leq0$. For $\alpha_2=0$, (\ref{hyp-V}) and $R_{uu}=-\Delta_hV$ give $R_{uu}=(\kappa/X_V)T^{\rm YM}_{uu}$. (c) $X_V>0>X_h$ requires $\alpha_3k>0$ and $n\geq3$. Then the Gauss--Bonnet terms in (\ref{XhXV}) are non-negative (the one in $X_h$ vanishes for $n=3$), so $X_h<0$ requires $Y<0$. (d) Eq.~(\ref{hyp-h}) with $t_h=0$ and $k\neq0$ gives $X_h=1+2\kappa k\,[(n-1)(n\alpha_1+\alpha_2)+(n-2)(n-3)\alpha_3]=0$.
\end{proof}

The obstruction is therefore not an artefact of Einstein's theory. Evading it costs either anti-gravitating waves, or a negative Gauss--Bonnet coupling together with an inverted effective Planck mass ($Y<0$), or the absence of any supporting matter (case (d), which includes the hyperbolic solutions of \cite{heefer}). The last two require transverse curvatures $|k|\gtrsim1/(\kappa|\alpha_i|)$, that is, curvatures at which the higher-curvature terms are as large as the Einstein term. On a hyperbolic transverse space, every bounded harmonic profile has bounded energy density, by the Cheng--Yau gradient estimate \cite{cheng-yau}; this supplies many regular gauge profiles, but a coupled wave also requires Eq.~(\ref{hyp-V}) to be solvable. In particular, if $\alpha_2=0$ and $X_V=0$, that equation forces $T^{\rm YM}_{uu}=0$ and hence $F=0$. On branch (d), $X_V=4\kappa\alpha_3(n-2)k$, so this obstruction occurs for $\alpha_2=0$ whenever $\alpha_3=0$ or $n=2$.

\subsection{Higher-order gauge theories}\label{sec:gauge-univ}

The gauge sector is equally insensitive to corrections. Kuchynka \cite{kuchynka} proved that null Yang--Mills fields solve every gauge theory whose Lagrangian is an analytic function of finitely many gauge-invariant polynomials in the field strength and its dual, with the Yang--Mills term as its quadratic part. For the coupled gravitational equations we also assume $\mathcal L(0)=0$; a constant term can instead be absorbed into $\Lambda$. The symmetrized-trace non-abelian Born--Infeld prescription, with its constant subtracted, is an example \cite{tseytlin-bi}. The reason is simple. By Lemma~\ref{lem:null}, $F^a{}_\mu{}^\alpha F^b{}_{\alpha\nu}=-\lambda_\mu\lambda_\nu\,\partial A^a\cdot\partial A^b$ and $\lambda^\nu F^c_{\nu\beta}=0$. Hence every product of three or more field strengths with Lorentz indices contracted in a chain vanishes, whatever the color structure, and every scalar built from $F$ vanishes, including those that involve the Levi-Civita tensor, since $F^a\wedge F^b=0$. A product of two field strengths with two free Lorentz indices is proportional to $\lambda_\mu\lambda_\nu$ and hence symmetric, so it cannot contribute to the antisymmetric tensor $\partial\mathcal L/\partial F^a_{\mu\nu}$. This also eliminates the quadratic expressions obtained by differentiating cubic invariants. On these configurations, $\partial\mathcal L/\partial F^a_{\mu\nu}$ and $\partial\mathcal L/\partial g^{\mu\nu}$ therefore reduce to their values for the quadratic part. If that part is $-\frac14c_{ab}F^a_{\mu\nu}F^{b\,\mu\nu}$ with an invariant, non-degenerate $c_{ab}$, the field equations are the Yang--Mills equations, and the energy--momentum tensor is (\ref{Tnull}) with $\gamma_{ab}$ replaced by $c_{ab}$.

Combined with the almost universality of the metrics, the solutions of this paper are exact in any higher-curvature theory coupled to any such gauge theory. Only the coefficients change: $\kappa\gamma_{ab}\to\kappa c_{ab}$ and $x\to P(x)$. Theories with derivatives of $F$ require a separate analysis; for aligned null fields in suitable Kundt spacetimes, see \cite{kuchynka}. Coupled Einstein--Yang--Mills solutions immune to corrections, including derivative and mixed gravity--gauge terms under additional geometric conditions, were studied in \cite{kuchynka-eym}.

\section{Conclusions}\label{sec:concl}

Coleman's non-abelian plane waves and their gravitating versions form a very restricted class of exact solutions. With flat transverse space, Coleman's profile is singled out twice. It is the only one with bounded energy density (as noted in $D=4$ by G\"uven \cite{guv1}; see also \cite{fuster-vh,fuster}), and the only one for which the whole Riemann tensor is the color-contracted square of the Yang--Mills field strength. The latter is an exact double copy of a spacetime with matter, with a constant zeroth copy, in which the gauge field is the actual source of the curvature. Both statements reflect the homogeneity of the wave fronts. For a Coleman wave with non-commuting amplitudes, the field strength shares this homogeneity, but the potential does not, even up to gauge; colored test particles can detect the difference through color memory, while neutral ones do not. In AdS we constructed a closed-form family of non-abelian waves in every dimension $D\geq4$. For $V_0=0$, back-reaction bounds the static-observer energy density in $D=4$ and for purely radial gauge profiles in any dimension, but not for every higher-dimensional profile. The nonzero scalar curvature excludes the flat-space Riemann-square identity in every dimension; in $D=4$, even the Weyl parts are not generically proportional. On a complete spherical transverse space, every non-trivial Yang--Mills wave in our ansatz is singular; higher-curvature corrections to the gravitational equations do not remove this obstruction. A hyperbolic transverse space would avoid this, but in quadratic and Gauss--Bonnet gravity it requires anti-gravitating waves or curvatures at which the higher-curvature terms become comparable to the Einstein term. Because both the metrics and the null gauge fields are (almost) universal, all of these statements hold, with modified coefficients, in generic gravity and gauge theories.

Several questions remain open: the physical interpretation of the singular solutions as fields of null color sources and their gravitational field; regular waves on non-compact curved transverse spaces in a healthy theory; non-affine solutions of higher-derivative theories; and the energy and holographic interpretation of the AdS waves.


\end{document}